\documentclass[apj]{emulateapj}

\newcommand{\etal}{et al.}

\newcommand\chandra{{\it Chandra}}

\newcommand\xmm{{\it XMM-Newton}}

\def\cco{CXOU~J232327.9$+$584842}
\def\snr{Kes~79}
\def\psr{\rm{PSR J1852$+$0040}}
\def\src{\rm{CXOU~J185238.6$+$004020}}
\def\one{\rm{1E~1207.4$-$5209}}

\def\puppsr{\rm{PSR J0821$-$4300}}

\def\simlt{\mathrel{\hbox{\rlap{\hbox{\lower4pt\hbox{$\sim$}}}\hbox{$<$}}}}
\def\simgt{\mathrel{\hbox{\rlap{\hbox{\lower4pt\hbox{$\sim$}}}\hbox{$>$}}}}

\slugcomment{}

\shorttitle{Spin-down Measurement of PSR J1852$+$0040}
\shortauthors{Halpern \& Gotthelf}

\begin{document}

\title{Spin-down Measurement of PSR J1852$+$0040 in Kesteven 79: \\
Central Compact Objects as Anti-Magnetars}

\author{J. P. Halpern \& E. V. Gotthelf}

\affil{Columbia Astrophysics Laboratory, Columbia University,
New York, NY 10027}

\begin{abstract}
Using \xmm\ and \chandra, we achieved phase-connected
timing of the 105~ms X-ray pulsar \psr\ that
provides the first measurement of the spin-down rate of a member
of the class of Central Compact Objects (CCOs) in supernova
remnants.  We measure $\dot P = (8.68 \pm 0.09) \times 10^{-18}$,
and find no evidence for timing noise or variations in X-ray flux
over 4.8~yr.  In the dipole spin-down formalism,
this implies a surface magnetic field strength
$B_s = 3.1 \times 10^{10}$~G, the smallest ever measured for a
young neutron star, and consistent with being a fossil field.
In combination with upper limits on $B_s$ from other
CCO pulsars, this is strong evidence in favor of
the ``anti-magnetar'' explanation for their low luminosity
and lack of magnetospheric activity or synchrotron nebulae. 
While this dipole field is small, it can
prevent accretion of sufficient fall-back material
so that the observed X-ray luminosity of 
$L_x = 5.3\times 10^{33}(d/7.1\ {\rm kpc})^2$~erg~s$^{-1}$
must instead be residual cooling.
The spin-down luminosity of \psr,
$\dot E =  3.0 \times 10^{32}$~erg~s$^{-1}$,
is an order-of-magnitude smaller than $L_x$.
Fitting of the X-ray spectrum to two blackbodies
finds small emitting radii, $R_1 = 1.9$~km and $R_2 = 0.45$~km,
for components of $kT_1 = 0.30$~keV and $kT_2 = 0.52$~keV,
respectively.  Such small, hot regions are ubiquitous
among CCOs, and are not yet understood in the
context of the anti-magnetar picture because
anisotropic surface temperature is usually
attributed to the effects of strong magnetic fields.
\end{abstract}


\section {Introduction}

The class of relatively faint X-ray sources
known as central compact objects (CCOs) in supernova remnants
(SNRs) are apparently isolated neutron stars, which we define
here by their steady flux, predominantly thermal X-ray emission,
lack of optical or radio counterparts, and absence of a 
surrounding pulsar wind nebula (see reviews by
\citealt{pav04} and \citealt{del08}).
Table~\ref{ccos} lists basic data on the well-studied CCOs,
as well as candidates whose
qualifications are not well established.
Of the seven most secure members,
three are definitely pulsars with periods of
0.105, 0.112, and 0.424~s \citep{got05,got09,zav00}.  Until now,
no spin-down was detected from a CCO,
which is most simply interpreted as indicating a weak surface
dipole magnetic field $B_s$.  In two cases, \psr\ ($P = 0.105$~s)
and \one\ ($P = 0.424$~s),
the {\it upper limits} on $B_s$ from spin period measurements are
$1.5 \times 10^{11}$~G and $3.3 \times 10^{11}$~G, respectively
\citep{hal07,got07}, smaller than that of any other young
neutron star.  We also
found $B_s < 9.8 \times 10^{11}$~G from a pair of observations
of \puppsr, the 0.112~s pulsar in Puppis~A \citep{got09}.

Important
implications of these results are that the birth periods 
of the CCO pulsars are not significantly different from their
present values, and that their spin-down luminosities are,
and have always been, insufficient to generate significant
non-thermal magnetospheric emission or synchrotron nebulae.
A corollary is that the so-called
``characteristic age'' $\tau_c \equiv P/2\dot P$ that is used
to approximate the true age of pulsar has, for CCOs, no meaning,
being at least millions of years
for pulsars that are demonstrably in supernova remnants that
are only a few thousand years old.  It is also reasonable to
suppose that some isolated radio pulsars with weak magnetic
fields that have characteristic ages of millions of years
may actually be former CCOs that are only moderately aged.

Most of the properties of the CCOs can thus be explained
by an ``anti-magnetar'' model \citep{hal07,got08},
including the possibility that their weak magnetic fields
are causally related to their slow rotation periods at birth through
the turbulent dynamo \citep{tho93}
that generates the magnetic field.  (See \citealt{spr08}
for a review of possible mechanisms for the origin of
magnetic fields in neutron stars.)
While CCOs are inconspicuous relative to
ordinary young pulsars and magnetars, the fact that
they are found in SNRs in comparable numbers
to other classes of neutron stars implies that they must represent
a significant fraction of neutron star births.
Considering that the CCO in Cassiopeia~A is the youngest known neutron
star (330~yr), and postulating that a CCO in SN 1987A
explains why its pulsar has not yet been detected, argues that
anti-magnetars are potentially a populous class.

\begin{deluxetable*}{llcccrccl}[t]
\tabletypesize{\scriptsize}
\tablewidth{0pt}
\tablecaption{Central Compact Objects in Supernova Remnants}
\tablehead{
\colhead{CCO} & \colhead{SNR} & \colhead{Age} & \colhead{$d$} & \colhead{$P$} &
\colhead{$f_p$\tablenotemark{a}} & \colhead{$B_s$} & \colhead{$L_{x,\rm bol}$} & \colhead{References} \\
\colhead{} & \colhead{} & \colhead{(kyr)} & \colhead{(kpc)} & \colhead{(s)} & \colhead{(\%)} &
\colhead{($10^{11}$~G)} & \colhead{(erg~s$^{-1}$)}
}
\startdata
RX~J0822.0$-$4300       & Puppis~A         & 3.7  & 2.2   & 0.112      & 11     & $<9.8$  & $6.5 \times 10^{33}$ & 1,2 \\
CXOU~J085201.4$-$461753 & G266.1$-$1.2     & 1    & 1 & \dots\ & $<7$    &\dots\ & $2.5 \times 10^{32}$ & 3,4,5,6,7 \\
1E 1207.4$-$5209        & PKS~1209$-$51/52 & 7    & 2.2   & 0.424      & 9      & $<3.3$  & $2.5 \times 10^{33}$ & 8,9,10,11,12 \\
CXOU~J160103.1$-$513353 & G330.2$+$1.0     & $\simgt 3$ & 5 & \dots\ & $<40$ & \dots\ & $1.5 \times 10^{33}$ & 13,14 \\
1WGA~J1713.4$-$3949     & G347.3$-$0.5     & 1.6  & 1.3 & \dots\   & $<7$   &\dots\   & $\sim 1 \times 10^{33}$ & 7,15,16  \\
CXOU~J185238.6$+$004020 & Kes~79           & 7    & 7   & 0.105      & 64     & $0.31$  & $5.3 \times 10^{33}$ & 17,18,19,20 \\
CXOU~J232327.9$+$584842 & Cas~A            & 0.33 & 3.4 & \dots\   & $<12$  &\dots\   & $4.7 \times 10^{33}$ & 20,21,22,23,24 \\
\hline
XMMU~J172054.5$-$372652 & G350.1$-$0.3     & 0.9  & 4.5 & \dots\ & \dots\ & \dots\ &  $3.4 \times 10^{33}$ & 25 \\
XMMU~J173203.3$-$344518
& G353.6$-$0.7     & $\sim 27$  & 3.2 & \dots\ & \dots\ & \dots\ &  $1.0 \times 10^{34}$ & 26,27,28 \\
CXOU~J181852.0$-$150213 & G15.9$+$0.2      & $1-3$ & (8.5) & \dots\   &  \dots\   & \dots\ & $\sim 1 \times 10^{33}$ & 29
\enddata
\tablecomments{Above the line are seven well-established CCOs.
Below the line are three candidates.}
\tablenotetext{a}{Upper limits on pulsed fraction are for a search down to $P=12$~ms or smaller.}
\tablerefs{
(1) \citealt{hui06};
(2) \citealt{got09};
(3) \citealt{sla01};
(4) \citealt{kar02};
(5) \citealt{bam05};
(6) \citealt{iyu05};
(7) \citealt{del08};
(8) \citealt{zav00};
(9) \citealt{mer02a};
(10) \citealt{big03};
(11) \citealt{del04};
(12) \citealt{got07};
(13) \citealt{par06};
(14) \citealt{par09};
(15) \citealt{laz03};
(16) \citealt{cas04};
(17) \citealt{sew03};
(18) \citealt{got05};
(19) \citealt{hal07};
(20) this paper;
(21) \citealt{pav00};
(22) \citealt{cha01};
(23) \citealt{mer02b};
(24) \citealt{pav09};
(25) \citealt{gae08};
(26) \citealt{tia08};
(27) \citealt{ace09};
(28) \citealt{hal09};
(29) \citealt{rey06}.
}
\label{ccos}
\end{deluxetable*}

The compact X-ray source \src\ was discovered in the
center of the SNR Kes 79 by \citet{sew03}.
In previous papers, we reported the discovery of 105~ms
pulsations from that CCO, now named \psr\ \citep[][Paper 1]{got05},
and the first few observations that established only an
upper limit on its period derivative corresponding to
$B_s < 1.5 \times 10^{11}$~G \citep[][Paper 2]{hal07}.
Here, we present a dedicated series of
timing observations of \psr\ that constitute
the first definite measurement of
the spin-down rate of a CCO pulsar, confirming
its unusually small dipole magnetic field and
supporting the anti-magnetar model.  The plan that was
devised to obtain the needed series of time-constrained
observations is described in Section 2.
Analysis and results of the timing data are presented in
Section 2.1, and the high-quality X-ray spectrum 
and pulse profile that were obtained
from the summed observations are the subject of Section 2.2.
Discussion of the anti-magnetar model, and possible
explanations for the X-ray spectrum appear in Section 3,
followed by the conclusions in Section 4.

In this paper, we adopt a
distance of 7.1~kpc to Kes~79 from \ion{H}{1}
and OH absorption studies \citep{fra89,gre92}
updated using the Galactic rotation curve
of \citet{cas98}.  We also assume the dynamical
estimate of 5.4--7.5~kyr for the age of the SNR from
\citet{sun04}.

\section{X-ray Observations}

After the first four timing observations of \psr\ in 2004 and 2006 (Paper~2),
it was evident that the frequency derivative
$\dot f$ was too small to measure without
obtaining a phase-coherent series of observations spanning several years. 
The most economical way to measure $\dot f$ is to begin
with a logarithmically spaced sequence of observations that maintains
the absolute cycle count $\phi(t)/2\pi$
and measures the frequency $f$ with increasing accuracy,
until the small quadratic term in the phase ephemeris 
\begin{displaymath}
{\phi(t) \over 2\pi} = f(t-t_0) + {1 \over 2}\dot f (t-t_0)^2 + . . .
\end{displaymath}
begins to make a significant contribution to the phase.
In this case,
it was also deemed possible (Paper~2) that accretion from
fall-back material could contribute timing irregularities
known as torque noise, of
a magnitude comparable to the existing upper limits on
dipole spin-down.  This made it all the more important to
obtain a well-sampled ephemeris that could test for such effects.

\begin{deluxetable*}{llrccccc}[t]
\tabletypesize{\scriptsize}
\tablewidth{0pt}
\tablecaption{Log of X-ray Timing Observations of \psr }
\tablehead{
\colhead{Mission} & \colhead{Instr/Mode} & \colhead{ObsID/Seq\#} & \colhead{Date} &
\colhead{Exposure} & \colhead{Start Epoch} & \colhead{Period\tablenotemark{a}} & \colhead{$Z^2_3$} \\
\colhead{} & \colhead{} & \colhead{} & \colhead{(UT)} & \colhead{(ks)} & \colhead{(MJD)} &
\colhead{(ms)} & \colhead{}        
}
\startdata
{\it XMM} & EPIC-pn/SW  & 0204970201   & 2004 Oct 18 & 30.6 & 53296.001 & 104.912638(39) & 121.7 \\
{\it XMM} & EPIC-pn/SW  & 0204970301   & 2004 Oct 23 & 30.5 & 53301.984 & 104.912612(55) & \phantom{1}77.2 \\
{\it XMM} & EPIC-pn/SW  & 0400390201   & 2006 Oct 08 & 29.7 & 54016.245 & 104.912610(47) & \phantom{1}92.4 \\
\chandra\ & ACIS-S3/CC  & 6676/500630  & 2006 Nov 23 & 32.2 & 54062.256 & 104.912592(40) & \phantom{1}94.0 \\
{\it XMM} & EPIC-pn/SW  & 0400390301   & 2007 Mar 20 & 30.5 & 54179.878 & 104.912600(43) & 107.5 \\
\chandra\ & ACIS-S3/CC  & 9101/500964  & 2007 Nov 12 & 33.1 & 54426.674 & 104.912615(40) & \phantom{1}95.8 \\
\chandra\ & ACIS-S3/CC  & 9102/500965  & 2008 Jun 16 & 31.2 & 54628.080 & 104.912593(40) & 115.5 \\
{\it XMM} & EPIC-pn/SW  & 0550670201   & 2008 Sep 19 & 21.2 & 54728.758 & 104.912563(84) & \phantom{1}61.7 \\
{\it XMM} & EPIC-pn/SW  & 0550670301   & 2008 Sep 21 & 31.0 & 54730.066 & 104.912594(46) & \phantom{1}84.2 \\
{\it XMM} & EPIC-pn/SW  & 0550670401   & 2008 Sep 23 & 34.8 & 54732.079 & 104.912573(42) & \phantom{1}80.3 \\
{\it XMM} & EPIC-pn/SW  & 0550670501   & 2008 Sep 29 & 33.0 & 54738.016 & 104.912596(40) & \phantom{1}97.4 \\
{\it XMM} & EPIC-pn/SW  & 0550670601   & 2008 Oct 10 & 36.0 & 54750.006 & 104.912641(40) & \phantom{1}90.0 \\
\chandra\ & ACIS-S3/CC  & 9823/501015  & 2008 Nov 21 & 30.1 & 54791.803 & 104.912645(55) & \phantom{1}69.8 \\
\chandra\ & ACIS-S3/CC  & 9824/501016  & 2009 Feb 20 & 29.6 & 54882.483 & 104.912591(53) & \phantom{1}80.4 \\
{\it XMM} & EPIC-pn/SW  & 0550671001   & 2009 Mar 16 & 27.0 & 54906.255 & 104.912610(48) & \phantom{1}97.9 \\
{\it XMM} & EPIC-pn/SW  & 0550670901   & 2009 Mar 17 & 26.0 & 54907.609 & 104.912592(60) & \phantom{1}82.9 \\
{\it XMM} & EPIC-pn/SW  & 0550671201   & 2009 Mar 23 & 27.3 & 54913.582 & 104.912616(46) & 104.8 \\
{\it XMM} & EPIC-pn/SW  & 0550671101   & 2009 Mar 25 & 19.9 & 54915.654 & 104.912544(85) & \phantom{1}75.8 \\
{\it XMM} & EPIC-pn/SW  & 0550671301   & 2009 Apr 04 & 26.0 & 54925.539 & 104.912590(52) & \phantom{1}97.9 \\
{\it XMM} & EPIC-pn/SW  & 0550671901   & 2009 Apr 10 & 30.5 & 54931.535 & 104.912620(48) & \phantom{1}82.9 \\
{\it XMM} & EPIC-pn/SW  & 0550671801   & 2009 Apr 22 & 28.0 & 54943.897 & 104.912621(69) & \phantom{1}67.4 \\
\chandra\ & ACIS-S3/CC  & 10128/501055 & 2009 Jun 02 & 33.2 & 54984.888 & 104.912596(37) & 117.9 \\
\chandra\ & ACIS-S3/CC  & 10129/501056 & 2009 Jul 29 & 32.2 & 55041.231 & 104.912633(40) & 105.6
\enddata									 
\tablenotetext{a}{Barycentric period derived from a $Z^2_3$ test.
The \cite{lea83} uncertainty on the last digits is in parentheses.}
\label{logtable}								 
\end{deluxetable*}

Unfortunately, maintaining cycle count would require more observations
classified as time-constrained than the \chandra\ {\it X-ray Observatory}
allocates to any one project, while for
\xmm, semiannual visibility windows for this source are only 40 days long,
not wide enough to securely bridge over the intervening 5 month gaps.
It took until 2008 to implement a strategy that uses the two satellites
in a coordinated fashion, filling two \xmm\
visibility windows with six observations each of variable
spacing, while requesting pairs of \chandra\ observations to
bridge the gaps between and after the two
\xmm\ windows.  In this manner, a phase-coherent timing solution
spanning 1.7~yr could be achieved, which could
also be extrapolated
backward to incorporate the earlier observations.
All but one of the approved observations in 2008--2009
were performed as planned.  Due to a loss of contact with the
\xmm\ spacecraft, the last observation of 2008 was rescheduled
to 2009, but this did not compromise the success of the program.  We were
thus able to obtain a fully coherent timing solution incorporating
all of the observations listed in Table~\ref{logtable}, including
the earliest ones, spanning 4.8~yr in total.

All of the \xmm\ observations used the pn detector of the
European Photon Imaging Camera (EPIC-pn) in ``small window'' (SW) mode
to achieve 5.7~ms time resolution, and an absolute uncertainty
of $\approx 3$~ms on the arrival time of any photon.
We processed all EPIC data using the emchain and
epchain scripts under Science Analysis System (SAS) version
xmmsas\_20060628\_1801-7.0.0.  The leap second at the end of 2008
was inserted manually.
Simultaneous data were acquired with the EPIC~MOS cameras, operated in
``full frame'' mode.  Although not useful for timing purposes because
of the 2.7~s readout, the location of the source at the center of the
on-axis MOS CCDs allows a better background measurement to test for flux
variability, an important indicator of accretion, than the EPIC-pn SW
mode. 

The \chandra\ observations used the Advanced Camera for Imaging and
Spectroscopy (ACIS) in continuous-clocking (CC) mode to provide time
resolution of 2.85~ms. 
This study uses data processed by the pipeline
software revisions v7.6.9--v8.0.  Reduction and analysis used
the standard software package CIAO (v3.4) and CALDB (v3.4.2).  The
photon arrival times in CC mode are adjusted in the standard
processing to account for the known position of the pulsar, spacecraft
dither, and SIM offset.
Absolute accuracy of the time assignment in
\chandra\ CC-mode is limited by the uncertainty in the position
of the pulsar, which was determined in Paper~2 from an ACIS image.
The typical accuracy of $\approx 1$ pixel then corresponds to an
uncertainty of $\approx 3$~ms, which is similar to the \xmm\
accuracy, and is 0.03 rotations in the case of \psr.  We will show that the
measured dispersion in pulse arrival times is comparable to this
uncertainty.

\subsection{Results of Timing Analysis}

For each observation in Table~\ref{logtable} we transformed the photon
arrival times to Barycentric Dynamical Time (TDB) using the 
pulsar coordinates determined in Paper~2 and reproduced in
Table~\ref{ephemeris}. 
Diffuse SNR emission is a significant source of background.
To maximize the pulsar signal-to-noise ratio, we used a source
extraction radius of $12^{\prime\prime}$ for the \xmm\ observations,
and five columns ($2.\!^{\prime\prime}4$) for the \chandra\ CC-mode data.
An energy cut of
$1-5$~keV was found to maximize pulsed power.
The pulse profile and value of the period in each
observation was derived from a $Z^2_3$ periodogram \citep{buc83},
a choice of harmonics that was found to minimize the uncertainties.
The resulting profiles were cross-correlated, shifted, and summed to 
create a master pulse profile template.  Individual profiles were
then cross correlated with the template to determine the time of
arrival (TOA) at each epoch. 

\begin{figure}[b]
\centerline{
\hfill
\includegraphics[scale=0.36,angle=270]{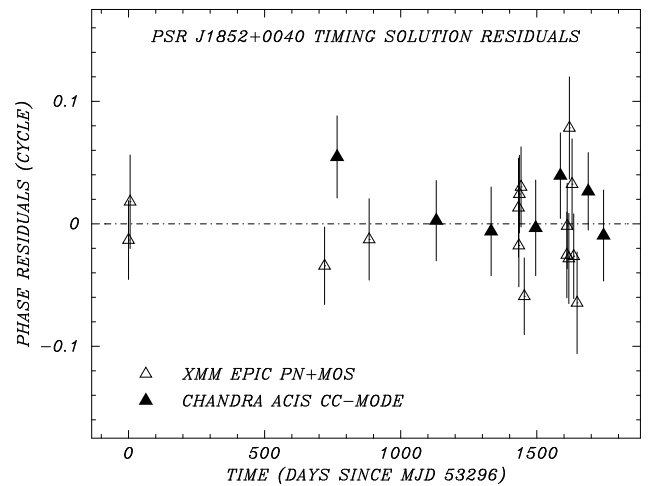}
\hfill
}
\caption{
Pulse phase residuals for the 23 timing observations of \psr\ 
from the best fitting ephemeris of Table~\ref{ephemeris}.
Error bars are 1 $\sigma$.
}
\label{phaseplot}
\end{figure}

\begin{figure}
\centerline{
\hfill
\includegraphics[scale=0.35,angle=270]{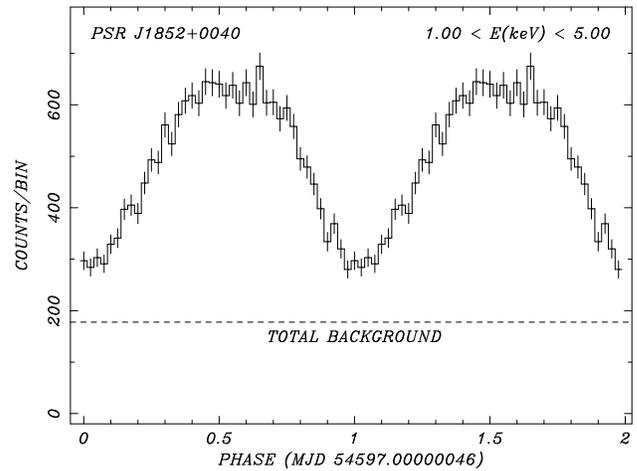}
\hfill
}
\caption{
Summed pulse profile from the 16 \xmm\ observations of \psr\
listed in Table~\ref{logtable}, folded using the ephemeris of
Table~\ref{ephemeris}. Phase zero corresponds to the TDB
epoch in Table~\ref{ephemeris}.  The pulsed fraction
after correcting for background is $f_p = 64\pm2\%$.
}
\vspace{-0.1in}
\label{lightcurve}
\end{figure}

Starting with the dense series of \xmm\ observations in
2008 September, the TOAs were iteratively fitted using the
TEMPO\footnote{http://www.atnf.csiro.au/research/pulsar/tempo}
software.  We fitted the three observations from
September 19--23 to a linear ephemeris, and added TOAs one at a time 
using a quadratic ephemeris, finding that the new TOA would match
to $<0.1$ cycles the predicted phase derived from the
previous set.  After adding the final observation to the
ephemeris, we then worked backward in time from 2008 September
to 2004 October until all 23 observations had been included.
The quadratic term contributes $-8.7$ cycles of
rotation over the 4.8~yr span of the ephemeris,
which yields the small uncertainty in $\dot P$ discussed below.
The phases and errors were determined by cross-correlating
with a final iterated template of co-added profiles produced from
the best fitting ephemeris given in Table~\ref{ephemeris}.
Figure~\ref{phaseplot} shows the residuals of the individual
observations from the best fit, which demonstrates the validity of the
solution.  The weighted rms of the phase residuals
is 3.4~ms, or 0.032 pulse cycles,
which is comparable to the individual measurement errors.
There is no evidence of timing noise or higher derivatives
in the residuals.  The variation of $Z^2_3$ (pulsed power) among
the observations listed in Table~\ref{logtable} is also
consistent with statistical expectations.
Figure~\ref{lightcurve} shows the summed pulse profile from
the 16 \xmm\ observations, for which it is possible to measure
a reasonably accurate background (unlike the \chandra\ CC-mode data).

The sensitivity of the results to $\dot P $ and $B_s$
can be estimated analytically.
If a pulsar is spinning down smoothly,
then a coherent timing solution spanning time $T$ will have
an uncertainty in frequency of
$\delta f = \delta P/P^2 \approx 0.1/T$, and will
be sensitive to a period derivative of
$\dot P_{\rm \min} = \delta \dot P \approx 0.2(P/T)^2$.
This in turn will
measure $B_s = 3.2 \times 10^{19}\sqrt{P\dot P}$
with fractional uncertainty
\begin{displaymath}
{\delta B_s \over B_s} \ \approx \ 1.0 \times 10^{38}\ {P^3 \over B_s^2\ T^2}
\end{displaymath}
\begin{displaymath}
\ = \ 0.05 \left(B_s \over 10^{10}\,{\rm G}\right)^{-2}
\left(T \over 4.8\ {\rm yr}\right)^{-2}
\left(P \over 0.105\ {\rm s}\right)^3.
\end{displaymath}
A coherent ephemeris for \psr\ spanning 4.8~yr has an uncertainty
on $\dot P$ of $\delta \dot P \approx 1 \times 10^{-19}$, corresponding
to a $5\sigma$ detection limit of $B_s = 5 \times 10^9$~G.
In comparison, the measured values $P=0.104912611147(4)$~s
and $\dot P = 8.68(9) \times 10^{-18}$
have fitted uncertainties consistent with these analytic estimates,
and imply in the dipole spindown formalism
a surface magnetic field strength $B_s = 3.1 \times 10^{10}$~G,
a spin-down luminosity $\dot E =
-I\Omega\dot\Omega = 4\pi^2I\dot P/P^3 = 3.0 \times
10^{32}$~erg~s$^{-1}$, and characteristic age
$\tau_c \equiv P/2\dot P = 192$~Myr, all with
negligible statistical error.

\begin{deluxetable}{ll}
\tablewidth{0pt}
\tablecaption{Spin Parameters of \psr }
\tablehead{
\colhead{Parameter} & \colhead{Value}
}
\startdata
Right ascension, R.A. (J2000)\tablenotemark{a} & $18^{\rm h}52^{\rm m}38^{\rm s}\!.57$ \\
Declination, Decl. (J2000)\tablenotemark{a}    & $+00^{\circ}40^{\prime}19^{\prime\prime}\!.8$ \\
Epoch (MJD TDB)\tablenotemark{b}               & 54597.00000046 \\
Spin period, $P$                               & 0.104912611147(4)~s \\
Period derivative, $\dot P$                    & $(8.68 \pm 0.09) \times 10^{-18}$ \\
Valid range of dates (MJD)                     & 53296 -- 55041 \\
Surface dipole magnetic field, $B_s$           & $3.1 \times 10^{10}$~G \\
Spin-down luminosity, $\dot E$                 & $3.0 \times 10^{32}$~erg~s$^{-1}$ \\
Characteristic age, $\tau_c$                   & 192~Myr
\enddata
\tablenotetext{a}{Measured from \chandra\ ACIS-I ObsID 1982 (Paper~1).
Typical \chandra\ ACIS coordinate uncertainty is $0\farcs6$.}
\tablenotetext{b}{Epoch of ephemeris corresponds to phase zero in
Figure~\ref{lightcurve}.}
\label{ephemeris}
\end{deluxetable}

\subsection{Flux and Spectral Analysis}

\begin{deluxetable*}{lcccc}[b]
\tablewidth{0pt}
\tablecaption{\xmm\ Spectral Fits to \psr }
\tablehead{
\colhead{Parameter}  & \colhead{BB} & \colhead{BB+BB} & \colhead{NSA} & \colhead{Comp BB}
}
\startdata
$N_{\rm H}$~($10^{22}$ cm$^{-2}$)         & $1.32 \pm 0.05$        & $1.82^{+0.23}_{-0.18}$  & $1.67^{+0.06}_{-0.07}$ & $1.53^{+0.09}_{-0.08}$ \\
$kT_1$ (keV)                              & $0.460 \pm 0.007$      & $0.30\pm0.05$ & $0.287 \pm 0.007$  & $0.396 \pm 0.014$    \\
$R_1$ (km)                                & $0.72\,d_{7.1}$        & $1.9\,d_{7.1}$  & 13.06\tablenotemark{a}  &  $0.93\,d_{7.1}$  \\
$kT_2$ (keV)                              & . . .                  & $0.52\pm0.03$   & . . . & . . .        \\
$R_2$ (km)                                & . . .                  & $0.45\,d_{7.1}$  & . . . & . . .       \\
$\Gamma$    & . . . & . . .  & . . .   & $4.80^{+0.34}_{-0.25}$ \\ 
$d$ (kpc)                                 & . . . & . . . &  $23.7^{+4.0}_{-2.6}$ & . . .  \\
$F_x(0.5-10\ {\rm keV}$)\tablenotemark{b} & $2.0 \times 10^{-13}$  & $2.0 \times 10^{-13}$ & $2.0 \times 10^{-13}$  & $2.0 \times 10^{-13}$  \\
$L_{\rm bol}$ (erg s$^{-1})$\tablenotemark{c} & $3.0 \times 10^{33}\,d_{7.1}^2$ & $5.3 \times 10^{33}\,d_{7.1}^2$ & . . .  & $3.6 \times 10^{33}\,d_{7.1}^2$ \\
$\chi^2_{\nu}(\nu)$                       & 1.38(204)  & 1.07(201) & 1.10(204) & 1.12(203)
\enddata
\tablecomments{Spectrum fitted over $0.7-7$~keV. Errors are 90\% confidence for one interesting parameter
($\chi^2 = \chi^2_{\rm min} + 2.7$).}
\tablenotetext{a}{Fixed parameter $R^{\infty}=13.06$~km corresponding to $R=10$~km with $M = 1.4\,M_{\odot}$.}
\tablenotetext{b}{Absorbed flux in units of erg cm$^{-2}$ s$^{-1}$.
Average of EPIC pn and MOS.}
\tablenotetext{c}{Unabsorbed, bolometric luminosity assuming $d = 7.1$~kpc.
Average of EPIC pn and MOS.}
\label{spectable}
\end{deluxetable*}

\begin{figure}[b]
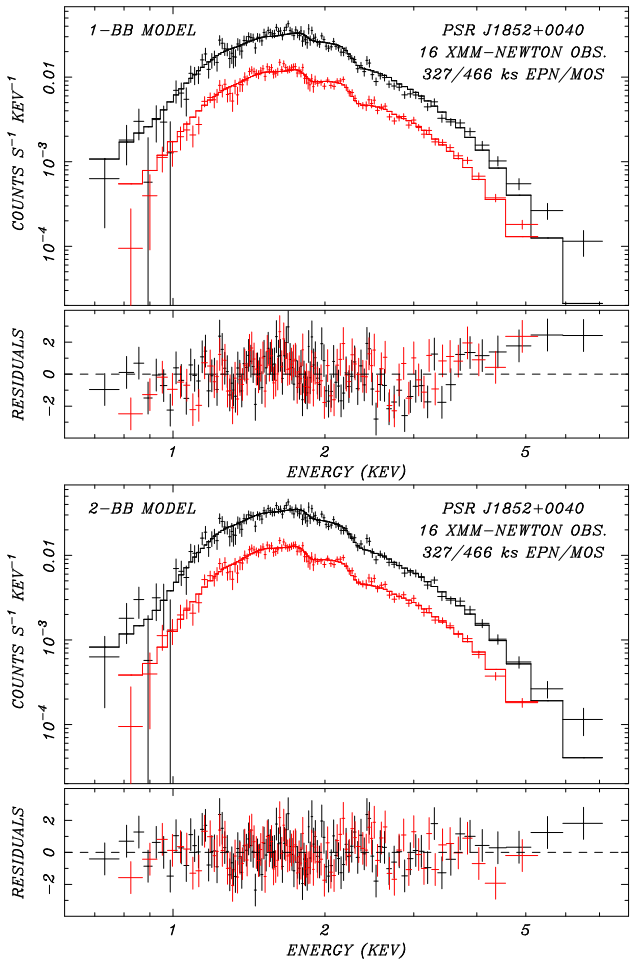

\centerline{
\includegraphics[scale=0.37,angle=270]{f4a.eps}
}
\centerline{
\includegraphics[scale=0.37,angle=270]{f4b.eps}
}
\caption{
Spectrum of the 16 summed \xmm\ observations of \psr.
The EPIC pn spectrum is black, while the average
of the MOS1 and MOS2 spectra is red.
Top: Fit to a single blackbody model.
Bottom: Fit to a double blackbody model. 
The parameters of the fits are
given in Table~\ref{spectable}.
}
\label{spectrum}
\end{figure}

In Papers 1 and 2 we presented several observations of \psr\ that were
consistent with steady flux and a blackbody spectrum of
$kT \approx 0.46$~keV.  The luminosity of $3.4 \times 10^{33}\,d^2_{7.1}$
erg~s$^{-1}$ corresponded to a blackbody radius of only
$\approx 0.8\,d_{7.1}$~km.  This is typical result for a CCO,
and indicates a concentrated hot spot that remains to be
understood.  Using the methods described in Paper 1,
we now combine all 16 \xmm\ observations of
\psr\ into one spectrum in order to search for deviations from a
blackbody that might indicate temperature variations on the surface,
or cyclotron lines.  EPIC pn and MOS spectra were fitted jointly.
The accumulated exposure allows us
to fit the spectrum over $0.7-7$~keV, an increase in coverage
over the $1-5$~keV range used in the previous papers.
While a single blackbody can still be fitted with a
temperature of 0.46~keV, the high signal-to-noise ratio and
expanded energy range of the summed spectrum reveal
systematic deviations from the fit; the reduced $\chi^2_{\nu} = 1.38$
for 240 degrees of freedom
is unacceptable (see Figure~\ref{spectrum} and Table~\ref{spectable}).
Similar to the results from other CCOs, we find that
a fit to two blackbodies is significantly improved ($\chi^2_{\nu} = 1.07$),
with temperatures of $kT_1 = 0.30$~keV and $kT_2 = 0.52$~keV,
while the corresponding radii $R_1 = 1.9\,d_{7.1}$~km and
$R_2 = 0.45\,d_{7.1}$~km still cover only a small fraction of the surface.
Here we define radius using the phase-averaged luminosity
$L_{1,2} = 4\pi R_{1,2}^2\sigma T_{1,2}^4$ regardless of the unknown
emission geometry, which could be, for example, a concentric annulus.
Adoption of the two-blackbody fit results in a higher bolometric
luminosity than the single blackbody,
$5.3 \times 10^{33}\,d_{7.1}^2$ erg~s$^{-1}$ vs.
$3.0 \times 10^{33}\,d_{7.1}^2$ erg~s$^{-1}$.
We don't combine the \chandra\
spectra here for an independent fit because of the increased
background and other uncertainties involved in analyzing
CC-mode data for this faint source.

\begin{figure}[b]
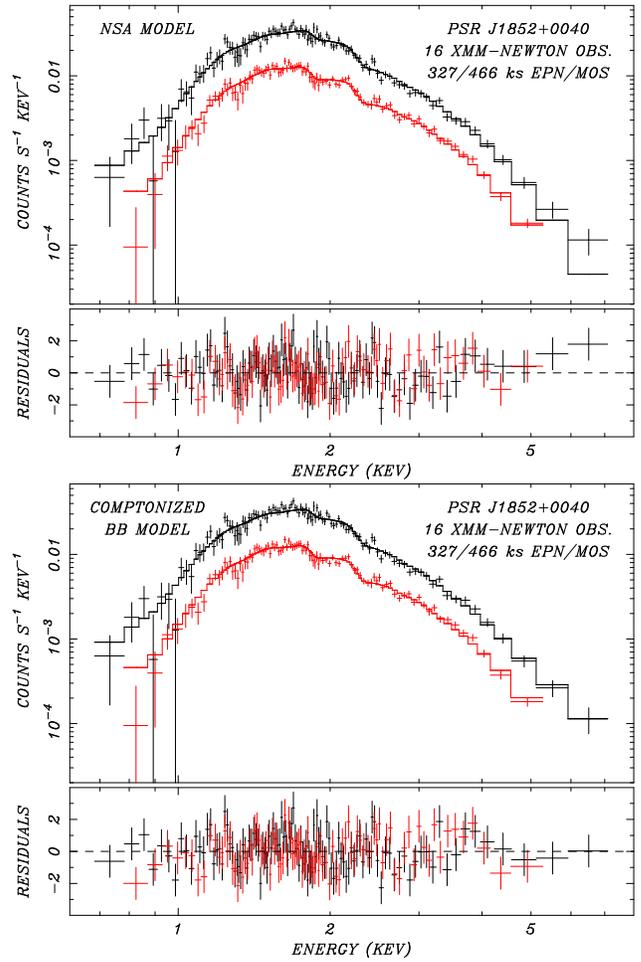

\centerline{
\includegraphics[scale=0.37,angle=270]{f5a.eps}
}
\centerline{
\includegraphics[scale=0.37,angle=270]{f5b.eps}
}
\caption{
Spectrum of the 16 summed \xmm\ observations of \psr.
The EPIC pn spectrum is black, while the average
of the MOS1 and MOS2 spectra is red.
Top: Fit to a nonmagnetic neutron star hydrogen atmosphere model.
Bottom: Fit to a Comptonized blackbody model. 
The parameters of the fits are
given in Table~\ref{spectable}.
}
\label{spectrum2}
\end{figure}

Additional spectral models that were explored include the
nonmagnetic neutron star hydrogen atmosphere of \citet{zav96},
and a simplified Comptonized blackbody, as described by
\citet{hal08} for application to anomalous X-ray pulsars.
Each of these models fits nearly as well as the two-blackbody model
because they can eliminate the residual excess at high energy
that is left by a single blackbody fit
(see Figure~\ref{spectrum2} and Table~\ref{spectable}).
For the neutron star atmosphere (NSA), we fixed the parameters
$M=1.4\,M_{\odot}$ and $R^{\infty}=13.06$~km (corresponding to $R=10$~km),
treating the effective temperature and distance as free parameters.
As expected, the fitted temperature is smaller than those of
the blackbody models, but the fitted distance of 23.7~kpc is
a factor of 3.3 higher than the known value.  This indicates
that in the atmosphere model only $\sim 10\%$ of the surface
is emitting.  We are not motivated to search for an atmosphere
model that fits to the full surface area, as was done for
the Cas~A CCO by \citet{ho09}, because the larged pulsed
fraction of \psr\ clearly requires a small hot spot.
The same interpretation attaches to the Comptonized
blackbody model, in which the inferred blackbody radius is only 
$0.93\,d_{7.1}$~km.

\begin{figure}
\centerline{
\hfill
\includegraphics[scale=0.47,angle=0]{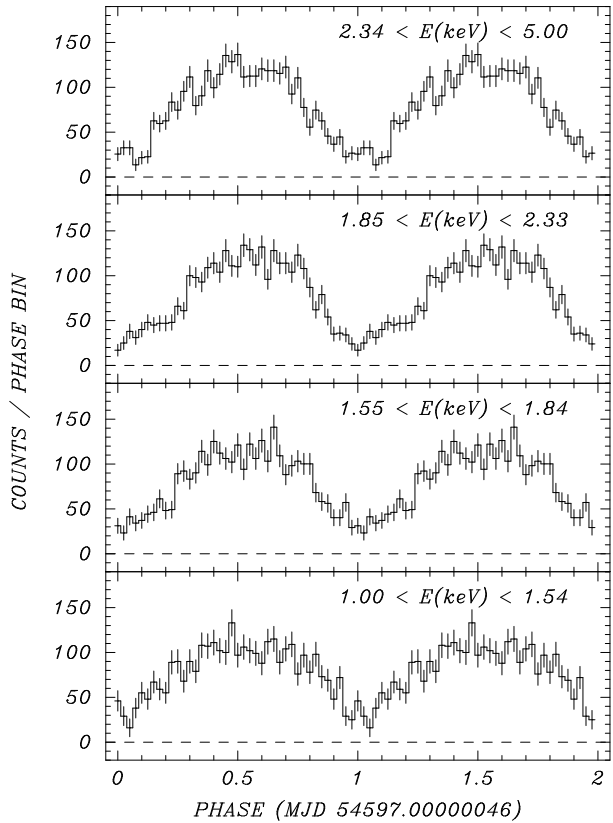}
\hfill
}
\caption{
Energy-resolved pulse profiles from the summed
16 \xmm\ observations of \psr\
of Figure~\ref{lightcurve} decomposed into
four energy bands containing equal numbers of photons.
Background has been subtracted.
The pulse shape does not differ significantly as a function of energy.
}
\label{fourfold}
\end{figure}

The large intrinsic pulsed fraction, $f_p = 64\pm2\%$ 
(defined as the fraction of counts above the minimum in the light-curve,
corrected for background; see Figure~\ref{lightcurve})
confirms that the emitting area must
be small and far from the rotation axis, while the line of sight also
makes a large angle to the rotation axis.
The pulse shape does not appear to be a function of energy
(Figure~\ref{fourfold}), which also suggests that the
emitting area is small.
The fitted column density from any model in Table~\ref{spectable}
agrees reasonably well with the
$N_{\rm H}$ derived from fits to the SNR spectrum. \citet{sun04}
find $N_{\rm H} = (1.54 -  1.78) \times 10^{22}$~cm$^{-2}$ from fitting
a variety of equilibrium and non-equilibrium ionization models to
{\it ASCA\/} and \chandra\ spectra of the SNR, while
\citet{gia09} find $N_{\rm H} = (1.52 \pm 0.02) \times 10^{22}$~cm$^{-2}$
from fitting a non-equilibrium ionization model
to the \xmm\ observations of 2004.

\begin{figure}[b]
\centerline{
\hfill
\includegraphics[scale=0.36,angle=270]{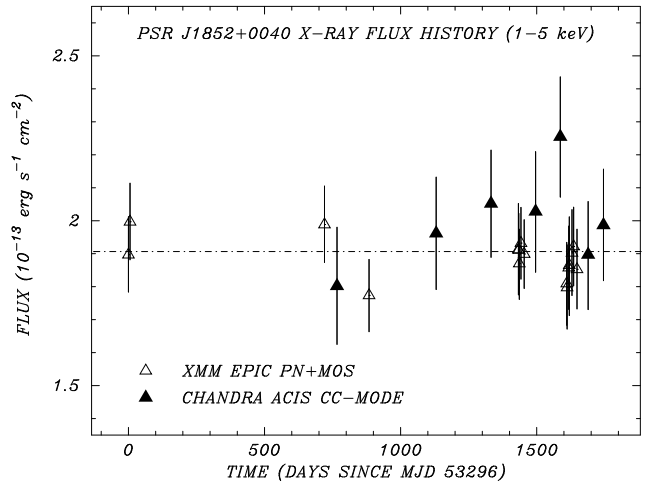}
\hfill
}
\caption{
Fluxes in the $1-5$ keV band for the 23 individual observations
of \psr, fitted to the two blackbody model given in Table~\ref{spectable}.
Errors bars are 1 $\sigma$.
The weighted mean flux is indicated by the dot-dashed line.
}
\label{fluxes}
\end{figure}

We then searched for any indication of variability
by extracting and fitting the spectrum of each individual observation,
fixing the spectral parameters (except for the total flux) to those
of the summed spectrum.  In order to minimize statistical and
systematic errors in this comparison, we restricted the fitted
energy range to $1-5$~keV for the individual spectra.
The resulting individual fluxes are shown
in Figure~\ref{fluxes}, where they are seen to be constant within errors.
The mean $1-5$~keV flux is $1.9 \times 10^{-13}$ erg~cm$^{-2}$ s$^{-1}$.
The typical $1 \sigma$ uncertainty in each observation is $\approx 10\%$,
and the rms dispersion among the 23 observations is $10\%$.

Unlike the unique case of \one, there is no evidence for cyclotron absorption
lines in the spectrum of \psr\ or in any other CCO.  \one\ has strong
absorption features at 0.7 and 1.4~keV \citep{san02,mer02a} that can
be attributed to the cyclotron fundamental energy $E_c$
and its first harmonic in a field of
$\approx 8 \times 10^{10}$~G \citep{big03}, 
according to the relation $E_c = 1.16(B/10^{11}\,{\rm G})/(1+z)$~keV,
where $z \approx 0.3$ is the gravitational redshift. The spectrum
of \psr\ can now be understood in terms of its weaker surface $B$-field.
For $B_s = 3.1 \times 10^{10}$~G, the cyclotron fundamental and first
harmonic are at 0.27~keV and 0.54~keV, below the region accessible
to X-ray spectroscopy (Figure~\ref{spectrum}), especially because of the larger
intervening column density to \psr.  The same explanation applied to other
well-observed CCOs suggests that they also have weaker surface $B$-fields
than \one.

\section{Discussion}

\subsection{CCOs as Anti-Magnetars}

The steady spin-down of \psr\ is consistent with magnetic braking of
an isolated neutron star with a weak magnetic field of
$B_s = 3.1 \times 10^{10}$~G.  In this case, the derived
characteristic age of 192~Myr
is not meaningful because the pulsar was
born spinning at its current period, $P = 0.105$~s, as was the case for
the other CCO pulsars, \one\ and \puppsr, which have
$P = 0.424$~s and $P = 0.112$~s, respectively.
A recent population analysis of radio pulsars by
\cite{fau06} favors such a wide distribution of birth periods (Gaussian
centered on $\sim 300$~ms, $\sigma \sim 150$~ms).

A $B$-field of this magnitude could be just the fossil field left
by flux conservation during the contraction of the core of the
progenitor star, as originally hypothesized for neutron stars
by \citet{wol64}.  Under this hypothesis, CCOs could simply be
those neutron stars in which no additional mechanism of $B$-field
amplification has been effective.  As stronger fields can be 
generated by a turbulent dynamo whose strength
depends on the rotation rate of the proto-neutron star \citep{tho93},
it is natural that pulsars born spinning slowly would have the weaker
$B$-fields.  The model of \citet{bon06} supports this, in particular
finding that the $B$-field of a slowly rotating neutron star should
be confined to small-scale regions, while a global dipole that 
would be responsible for spin-down is absent.

\begin{figure}
\centerline{
\hfill
\includegraphics[scale=0.43]{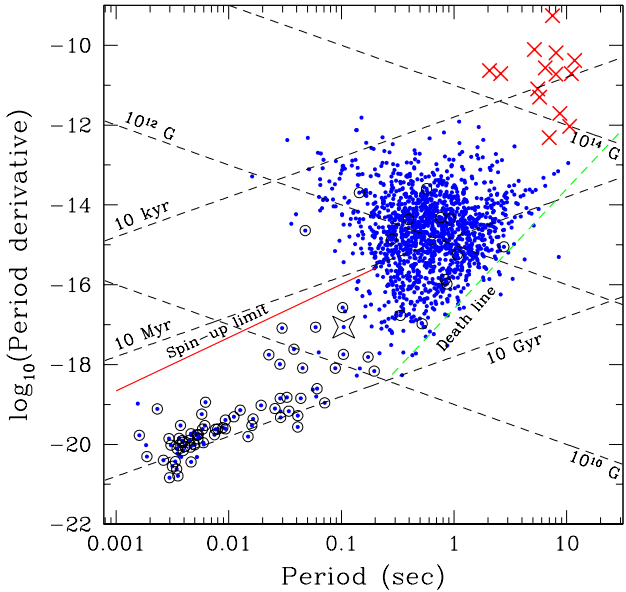}
\hfill
}
\vspace{-0.2in}
\caption{$P-\dot P$ diagram of isolated pulsars (dots),
binary radio pulsars (circled dots), and magnetars (crosses).
The location of \psr\ is marked by the star.
The radio pulsar death line $B/P^2 = 1.7 \times 10^{11}$ G~s$^{-2}$ of
\citet{bha92} is indicated.  The spin-up limit for recycled
pulsars corresponds to
$P({\rm ms}) = 1.9\,(B/10^9\,{\rm G})^{4/3}$ \citep{van87}.
}
\vspace{-0.1in}
\label{ppdot}
\end{figure}

\psr\ falls in a region of ($P,\dot P$) parameter space
(Figure~\ref{ppdot}) that is
devoid of ordinary (non-recycled) radio pulsars \citep{man05}.
It overlaps with recycled pulsars in this area.
The spin parameters of \psr\ are not beyond the empirical or
theoretical radio pulsar death lines of \citet{fau06} or \citet{che93},
respectively.
One possible explanation for the absence of radio emission from CCOs
is low-level accretion of SN debris for thousands or even millions
of years.  However, the sample of CCOs is not
yet large enough to know if they are intrinsically radio quiet.

Binary pulsars with similar spin parameters
as  \psr\ are thought to have been partly 
spun up by accretion.  It was also suggested by \citet{des95}
that most single pulsars in this region are recycled,
the birth rate of pulsars with
$B_s < 10^{11.5}$~G being too low in their estimation,
perhaps one in $\sim 5000$~yr.
To the contrary, \citet{har97}
concluded that it is not possible to distinguish ordinary pulsars
of $B_s = 10^{10.5-11.5}$ from recycled ones, and that there is
no reason to suppose that any such single pulsar is recycled.
The measurements of spin parameters of CCOs certainly increases
the inferred birth rate of neutron stars in this weak $B$-field regime,
a renewed warning not to assume that isolated radio pulsars
with similar spin parameters are recycled.
It should be expected that many
of the $\approx 52$ radio pulsars with $B_s < 10^{11}$~G
and $0.1<P<0.7$~s in Figure~\ref{ppdot} have characteristic ages
that are meaningless, being hundreds of millions of years.
They have not moved in the ($P,\dot P$) diagram,
and may be former CCOs whose SNRs dissipated only
$10^5-10^6$~yr ago.  Once the SNR has disappeared
and the neutron star has cooled, it is difficult to recognize
and classify such an ``orphan CCO'' using X-rays unless it was first
selected as a radio pulsar. Allowing for this observational handicap,
CCOs having weak $B$-fields may represent a channel of neutron star
birth that is as common as any other.

This outcome is counter in its details to the long debated
``injection'' hypothesis.
Using the ``pulsar current'' analysis proposed by \citet{phi81},
\citet{viv81} and \citet{nar87} suggested that a large fraction 
of pulsars are injected with $0.5 < P < 1.0$~s and
$\dot P > 1 \times 10^{-14}$ (i.e., with large magnetic field),
or that neutron stars do not turn on as radio pulsars until
their $\dot P$ decreases below a critical value of $\sim 3 \times 10^{-13}$.
This would require that they also cool rapidly in order not to be detected
as thermal X-ray sources in SNRs.  Later investigators
\citep[e.g.,][]{lyn85,lor93} argued that
this interpretation suffers from substantial statistical
uncertainties and selection effects
in pulsar surveys, and that injection of a distinct population
of pulsars is not necessary.
While previous authors dismissed the possibility
of injection of radio pulsars with weak $B$-fields \citep[e.g.,][]{sri84},
this appears to be exactly the origin of radio-quiet CCOs.
The related problem of the missing pulsars in empty-shell
SNRs \citep{got98} has largely been solved in recent years through
a variety of channels, as summarized by \citet{kas02}, including
the detection of radio-faint pulsars, pulsar wind nebulae
in X-rays, and now, pulsations from CCOs.

The spin-down power $\dot E = 3.0 \times 10^{32}$ erg~s$^{-1}$
of \psr\ is an order of magnitude smaller than its observed thermal
X-ray luminosity, $L_x \approx 5.3 \times 10^{33}\,d_{\rm 7.1}^2$ erg~s$^{-1}$,
which argues that the latter is mostly residual cooling.
Remaining questions presented by the X-ray properties of \psr\
and other CCOs are focussed on the details of their X-ray spectra
and pulse profiles, which require small heated areas.
In the case of \psr, the spectrum is more complex than
a single blackbody.  In the two blackbody model, for
example, the hotter component, of temperature $kT_2 = 0.52$~keV,
has an area of only $\sim 2.5\,d^2_{7.1}$~km$^2$. 
The canonical open-field-line polar cap has area
$A_{\rm pc} = 2\pi^2R^3/Pc \approx 1.1$~km$^2$, which
is comparable to the area of the hot spot, but polar cap
heating by any magnetospheric accelerator must be
negligible in this case, being much smaller than the
spin-down power.  Accretion may cover a wider area
and generate more luminosity, but that process is also
problematic for \psr, as discussed below.
In the remainder of this section we discuss alternative
hypotheses for the weak surface $B$-fields and thermal hot
spots of CCOs, involving internal magnetic fields,
anisotropic conduction, and accretion of fall-back material.
Although these theories have some attractive properties,
they are speculative, and none yet offers a self-consistent
explanation for all of the observed properties of CCOs.

\subsection{Localized Heating by Magnetic Field Decay?}

While the X-ray luminosity of \psr, $5 \times 10^{33}$ erg~s$^{-1}$,
is consistent with minimum neutron star cooling scenarios
\citep{pag04,pag07} for the age of \snr\ \citep[5.4--7.5 kyr,][]{sun04},
the high temperatures and small
surface areas fitted to the X-ray spectrum are difficult
to understand without invoking either localized heating on the surface
or strongly anisotropic conduction.
We recall that the X-ray spectra and luminosities
of CCOs are not very different from those of quiescent magnetars,
another recently recognized class,
being attributable largely to one or two surface thermal 
emission components \citep[e.g.,][]{hal05}.  It is difficult
to distinguish CCOs from quiescent magnetars without
timing data or evidence about variability \citep{hal09}.
Thus, it is tempting to hypothesize that the same magnetic field
decay that is thought to be responsible for localized crustal 
heating in a magnetar can be operating in CCOs.  However, in
the case of CCOs, such $B$-fields would need to have the same
$10^{14}-10^{15}$~G strengths as in magnetars to account
for continuous X-ray luminosities of $10^{33}-10^{34}$ erg~s$^{-1}$,
while having insignificant dipole moments that do not contribute to
spin-down.  One such configuration would be a small ``sunspot''
dipole on the rotational equator, to account for the large pulse
modulation, while the rest of the star contributes
dipole $B_s < 3 \times 10^{10}$~G in order not to exceed the
observed spin-down rate.  While we cannot rule out this
possibility, it would be remarkable if the magnetic field of a neutron
star could be created or evolve into a configuration with
such extreme contrast.  Also, the conspicuous lack of X-ray variability
from CCOs argues against invoking the same magnetic field strength and
heating mechanism that is held responsible for magnetars, with their
ubiquitous variability.

\subsection{Anisotropic Conduction?}

We next turn to the possibility that magnetic field confined entirely
beneath the surface can be responsible for the nonuniformity of
surface temperature.  While a strong $B$-field
is a favored ingredient of models for anisotropic conduction,
\psr\ with its exceptionally weak dipole field
would seem the least likely candidate for such an explanation.
Nevertheless, it is possible that a strong toroidal field can exist
under the surface, while only a weak external poloidal field
contributes to spin-down.  A toroidal field is expected to be
the initial configuration generated by differential rotation in 
the proto-neutron star dynamo \citep{tho93}.  The effects
on the surface temperature distribution of an internal
toroidal field were calculated by \cite{per06}, \citet{gep06},
and \citet{pag07}.
One of the effects of crustal toroidal field is to
insulate the magnetic equator from heat conduction, resulting
in warm spots at the poles.  It was even shown that
the warm regions can be of different sizes due
to the antisymmetry of the poloidal component
of the field, which is reminiscent of the asymmetric opposing
thermal regions in the Puppis~A CCO \citep{got09}.

In order to explain the large observed pulsed fraction of \psr,
the axis of toroidal magnetic field must make a large
angle with respect to the rotation axis of the neutron
star.  An orthogonal configuration was in fact
adopted by \citet{gep06} and \citet{pag07}.
Even if the toroidal field component is initially
parallel to the rotation axis,
a toroidal field will deform the neutron star
into a prolate shape, which tends toward
orthogonality to the rotation axis
as a minimum energy configuration \citep{bra09}.
The main problems with this mechanism are 1)
anisotropic conduction is unlikely to produce
a small hot region on \psr\ that covers only
$\sim 1\%$ of the surface area of the neutron star,
and 2) it doesn't explain why 
only one of the orthogonal poles is evidently hot.

Furthermore, to have a significant effect
on the heat transport, the crustal toroidal
field strength required in these models is $\sim 10^{15}$~G,
many orders of magnitude greater than the poloidal field
if the latter is measured by the spin-down.
Purely toroidal or poloidal fields are thought
to be unstable \citep{tay73,flo77}, although the toroidal
field may be stabilized by a poloidal field that is several
orders of magnitude smaller \citep{bra09}, so this may be
a viable configuration for a CCO.  On the other hand,
twisting and breaking of the crust by such a large toroidal
field is the basis of the magnetar model for soft gamma-ray
repeaters and anomalous X-ray pulsars \citep{tho02},
which also have large
external dipole fields as measured by their rapid spin-down.
In this picture, a CCO is a magnetar-in-waiting,
a scenario that \citet{pav09} found somewhat contrived,
as do we.  If {\it any} pulsar {\it could\/}
be an incipient magnetar, it isn't explained why CCOs
have especially weak surface fields compared to ordinary pulsars.

\subsection{Submergence of Magnetic Field by Hypercritical Accretion?}

For as long as the SN explosion mechanism has been studied,
it has been noted that a newly born neutron star may
accrete large amounts of fall-back material in the hours
to months after the SN.  This could be submerge
the initial magnetic field into the core \citep{gep99},
and the surface field could be essentially zero if the accreted
matter is not highly magnetized in a well-ordered fashion.
After the accretion stops, the submerged field will diffuse
back to the surface \citep{mus95}, but this could take hundreds
to millions of years \citep{gep99}.  In this picture, the CCOs
could be those neutron stars that suffered the most fall-back
accretion, while the normal pulsars with ages of a few thousand years
must have accreted  $<<0.01\,M_{\odot}$.  In the model
of \citet{mus95} with accretion of $\sim 10^{-5}\,M_{\odot}$,
the regrowth of the surface field is largely complete after $\sim 10^3$~yr,
so this would probably not explain the weak field of \psr.
But if $> 0.01\,M_{\odot}$ is accreted, then the
diffusion time could be millions of years.
\citet{che89} calculated that the neutron star in SN~1987A could
have accreted $\sim 0.1\,M_{\odot}$ of fallback material in the hours after the
SN explosion, aided by a reverse shock from the helium layer
of the progenitor.  If so, it may never emerge as a radio pulsar.
A more normal type II SN should accrete
$10^{-3}\,M_{\odot}$ or less, which may delay the emergence
of the magnetic field for tens of thousands of years,
a timescale applicable to explaining the properties of CCOs.

Finally, it should be considered that a thermoelectric instability
mechanism driven by the strong temperature gradient in the
outer crust \citep{bla83} could regenerate the magnetic field submerged
by accretion.  However, the growth of the field via this
mechanism could take $\sim 10^5$~yr.
CCOs could then be those neutron stars in which the thermoelectric
instability is the dominant mechanism of generating magnetic
field, perhaps because their initial rotation rates are slow.

The models discussed here have the effect of delaying the emergence
of a normal magnetic field for times ranging from a thousand
years to essentially forever.  They are difficult to
test using CCOs, for which it is impossible to
measure the braking index, which could indicate field amplification.
Furthermore, they don't help to explain
the small, hot surface areas seen in the X-ray spectra
of CCOs.  In order to achieve such a configuration,
the magnetic field would have to be submerged everywhere except
for one or two small regions, perhaps the magnetic poles of the
neutron star, which could remain hot via conduction from the
interior.

\subsection{Continuing Accretion from a Fall-Back Disk?}

We also consider ongoing accretion from a fall-back disk of supernova
debris as a possible source of anisotropic surface heating.
While variability is an indicator of accretion, we do not have
evidence of any variations of \psr\ at the 10\% level
among 23 observations spanning
4.8~yr, which would tend to eliminate accretion as a significant
contributor to its X-ray luminosity.
However, although accretion
is widely considered to be an inherently unstable process, 
it is not clear that variability of X-ray binaries or
AGNs can be extrapolated to accretion from a fossil
disk at rates of only $\sim 10^{-5}\,L_{\rm Edd}$.  Therefore, we also
explore the possibility of accretion, constrained only by the steady
spin-down rate of \psr, using the theory of propeller and accretion-disk
torques.

The spin parameters of \psr\ fall in a regime in which both
dipole braking and accretion torques are conceivably significant.
In Paper~2, we discussed the implications for $\dot P$ of \psr\ accreting
and spinning down in the propeller regime, deriving
\begin{displaymath}
\dot P \approx 2.2 \times 10^{-16}\,\mu_{28}^{8/7}\,\dot M_{13}^{3/7}
\left({M \over M_{\odot}}\right)^{-2/7}I_{45}^{-1}
\left({P \over 0.105\ {\rm s}}\right)
\end{displaymath}
\begin{displaymath}
\times \left(1- {P \over P_{\rm eq}}\right)
\end{displaymath}
for the propeller effect.
In this model, $\dot M$ is the rate of mass expelled,
which must be $> \dot m$, which is the accretion rate
onto the neutron star.  We can therefore assume 
$L_X = \eta\,\dot m\,c^2$ to calculate a lower limit
on $\dot M$ in the propeller accretion scenario.
Assuming efficiency $\eta \sim 0.1$, $\dot M > 5 \times 10^{13}$ g~s$^{-1}$
is required.  But if we were to adopt $\mu = B_s\,R^3$, where
$B_s = 3.1 \times 10^{10}$~G from assuming dipole spin-down,
then the observed $\dot P = 8.68 \times 10^{-18}$
allows a negligible $\dot M < 3 \times 10^{8}$ g~s$^{-1}$,
which contradicts the accretion assumption.

If instead we use the measured $\dot P$
as an upper limit on the propeller spindown rate, we can derive an
upper limit on the required $\mu$ in the accretion scenario.
Under these assumptions,
$\mu < 3.4 \times 10^{26}$ G~cm$^3$, or $B_s < 3.4 \times 10^9$~G.
However, this would marginally violate the conditions of the propeller
model, because for such a small magnetic field, the magnetospheric
radius is
\begin{displaymath}
r_m\ =\ 2.3 \times 10^7\,\mu_{26}^{4/7}\,\dot M^{-2/7}_{13}
\,\left({M \over M_{\odot}}\right)^{-1/7}=\ 2 .6 \times 10^7\ {\rm cm},
\end{displaymath}
which is smaller than the corotation radius 
\begin{displaymath}
r_{\rm co}\ =\ \left(G\,M\right)^{1/3}\left({P \over 2\pi}\right)^{2/3}
=\ 3.7 \times 10^7\ {\rm cm},
\end{displaymath}
and the pulsar would be rotating near its equilibrium period.
But there has not been enough time in the life of \snr\ for
a weakly accreting pulsar to come into equilibrium.
To find \psr\ in an equilibrium state would be a remarkable
coincidence because its spin-down timescale,
$P/\dot P \sim 4 \times 10^8$~yr, is orders of magnitude
greater than its age.
Therefore, it is more natural to conclude that
the small $\dot P$ of \psr\ is due purely to dipole spin-down,
and that it is not accreting.  Absence of accretion would
also argues against nonuniform surface composition
as a cause of the temperature variations.
Finally then, none of the possible
mechanisms discussed here for hot spots on CCOs is entirely
natural and self-consistent.

\section{Conclusions and Further Work}

Measurement of the spin-down rate of the 105~ms \psr\ in \snr\
was achieved using X-ray observations coordinated between
\xmm\ and \chandra.  The resulting phase-connected ephemeris
spanning 4.8~yr requires only the first derivatives of the
spin frequency, with no measurable timing noise superposed.
The straightforward interpretation of this
result is dipole spin-down due to a weak surface
magnetic field of $B_s = 3.1 \times 10^{10}$~G,
the smallest measured for any young neutron star.
While this is the first measurement of the spin-down rate of
a CCO, upper limits on the period derivatives of
\one\ and \puppsr, as well as spectral features in those two,
also indicate weak $B$-fields.  The
properties of the CCO class were loosely defined
in the past, but it is now evident that a weak $B$-field is the
physical reason that a large fraction
of the neutron stars were so named.  The body of evidence
supports the ``anti-magnetar'' model for the origin of
such CCOs: neutron stars born spinning ``slowly'' may as a
result be endowed with only weak magnetic fields.  \psr\
falls in a region of ($B,P$) space that overlaps with
moderately recycled pulsars.  Single radio pulsars with
similar spin parameters may therefore be former CCOs rather than
recycled, and they may be much younger than their characteristic ages.
X-ray observations of pulsars in this region may
find evidence of their relative youth via surface thermal emission.

Ongoing timing studies of the CCOs \one\ and \puppsr\
will able to measure their spin-down rates and can 
refine the correspondence between spectral features interpreted
as electron cyclotron lines and surface $B$-field.  Those CCOs such as
\psr, with no apparent absorption lines, may simply have weaker
$B$-fields than the others
by a factor $2-3$, so that the cyclotron resonance
is below the soft X-ray band.  As for CCOs that have not
yet been seen to pulse, there is no reason to suppose that they
are of a different physical class.  Rather, their surface temperature
distributions and/or viewing geometries may deliver only weak modulation
that happens to fall below the sensitivity of existing data,
while deeper observations may succeed in discovering their spins.
It would be of great interest to discover
pulsations from the CCO in Cas~A, the youngest known neutron
star, which in all respects 
is the prototype of the CCO class.  At an age of 330~yr, it
is an order of magnitude younger than the others,
while having similar spectral properties
\citep{pav09}.  In our model, its spin parameters
were fixed at birth, and should conform closely
to those of the older CCOs.  A recent \chandra\
observation of the Cas~A CCO did not detect pulsations
(see Appendix), but the High Resolution Camera (HRC) 
that was employed lacks energy resolution, a capability
that may be crucial in the same way that it was
for the \xmm\ discovery of the pulsar in Puppis~A \citep{got09}.

A remaining theoretical puzzle about CCOs is the origin
of their surface temperature anisotropies, in particular, the one
or two warm/hot regions that are smaller than the full neutron star
surface.  The high signal-to-noise spectrum accumulated from
\psr\ reveals similar temperature structure as the other CCOs.
The measured spin-down power of \psr\ is too small by an order of
magnitude to contribute to these excess emissions.
We considered whether accretion of fall-back material can
be responsible for this effect, but find it unlikely in the
case of \psr\ because of its slow, steady spin-down.  If other
CCOs are found to have similar spin-down properties,
the same arguments would apply to them.  Therefore, it is
important to measure the spin-down rate of \puppsr, the CCO
in Puppis~A, and investigate further its apparent phase-dependent emission
line at 0.8~keV \citep{got09}, which may be indicative of
accretion because it is in emission.

The small regions of high surface temperature are 
properties that CCOs share with magnetars, although
most magnetars are hotter and more luminous.
The total X-ray luminosities of CCOs are
consistent with slow cooling scenarios, and
there would be no need to hypothesize a
mechanism of magnetic field amplification beyond simple flux
freezing if it were not for their strongly anisotropic surface
temperature distributions.  Paradoxically, the
explanations that have been considered for hot spots invoke
strong magnetic fields just below the surface, which is exactly
the basis of the magnetar model.  In this sense, CCOs are
potentially magnetars waiting to emerge, or maybe ordinary
neutron stars in which the original field was submerged
by especially massive fall-back of supernova debris.
This is an unsatisfying
picture, if only because the contrast between
field strength in different regions must be a factor of $10^4$
or larger to affect the surface temperature distribution
while maintaining the slow spin-down and allowing 
soft X-ray cyclotron lines.
Resolving this essential mystery of the CCOs will undoubtedly
generate new insight into the physics of neutron stars.
\acknowledgements
We thank Fernando Camilo for supplying Figure 5.
This investigation is based on observations obtained with \xmm,
an ESA science mission with instruments and contributions directly funded by
ESA Member States and NASA, and \chandra.
The opportunity to propose a large project that
included coordinated observations between \xmm\ and \chandra\
was crucial for the success of this long-term effort. 
Financial support was provided by NASA through {\it XMM\/} grant
NNX08AX71G and {\it Chandra} Awards SAO GO8-9060X, SAO GO9-0058X,
and SAO GO9-0080X issued by the \chandra\ X-ray Observatory Center,
which is operated by the Smithsonian Astrophysical Observatory for
and on behalf of NASA under contract NAS8-03060.

\appendix

\section{Search for X-ray Pulsations from the Cas~A CCO}

In this Appendix we report on analysis of a recent \chandra\
observation designed to be the most sensitive yet to search
for pulsations from \cco, the CCO in Cas~A.  But first
we summarize the results of previous attempts.
\citet{mur02} searched a 50~ks \chandra\
HRC observation and identified several candidate periods,
but none was confirmed in a follow-up 50~ks observation \citep{ran02}.
\citet{pav09} found from a \chandra\ ACIS observation
taken in subarray mode $f_p(3\sigma)<16\%$ for $P > 0.7$~s.
That mode is limited by its sampling time of $0.34$~s.
ACIS in full-frame mode has a sampling time of 3.2~s, so
is only sensitive to periods $>6.4$~s. For the 1~Ms of
data taken in this mode \citep{hua04} we searched the range
$6.4 \le P \le 100$~s and $\dot f \le 1 \times 10^{-11}$ Hz~s$^{-1}$,
corresponding to $\tau_c \ge 330$~yr.  The resulting upper limit
is $f_p < 4\%$ at 99\% confidence.
Existing \xmm\ data on \cco\ are also limited in sensitivity and
time resolution, and do not cover the full range of periods spanned
by known CCO pulsars.  Using an effective exposure of 9.2~ks in the
EPIC pn taken in full-frame mode,
\citet{mer02b} reported an upper limit of $f_p < 13\%$ for
$P > 0.3$~s, and $f_p < 7\%$ for $P > 3$~s.  However, these
limits may have been underestimated by a factor
of $2-3$ \citep[see][]{pav09}.  Our own analysis sets
limits of $f_p<32\%$ for $0.146\,{\rm ms} < P \le 3$~s
and $f_p<17\%$ for $P > 3$~s from this observation.
Improvements
in sensitivity and time resolution could be obtained with a longer
\xmm\ observation using the EPIC pn in SW mode, as was employed
for the other CCO pulsars.

Recently, 487~ks of \chandra\ HRC timing data
were obtained over a period of 11 days in 2009 March,
of which 433~ks are in the public archive.
We searched the public data for pulsations.
A total of 11,486 counts were extracted from a circular aperture
of radius $1.\!^{\prime\prime}2$, of which 6\% are estimated to be
background.
We sampled ($f,\dot f$) parameter space using the $Z_1^2$ test, which is
an optimal one for a source in which the light curve, presumed to
arise from surface thermal emission, is expected to be weakly
modulated and quasi-sinusoidal.  We limited the search range
to $f \le 200$~Hz for $\dot f \le 5 \times 10^{-13}$~Hz~s$^{-1}$,
$f \le 100$~Hz for $\dot f \le 3 \times 10^{-12}$~Hz~s$^{-1}$,
and $f \le 10$~Hz for $\dot f \le 1 \times 10^{-10}$~Hz~s$^{-1}$,
oversampling by a factor of $\approx 3$ in each parameter.
This range covers the parameters of all the known CCO pulsars
and magnetars.
The largest values of $Z_1^2$ found were $\approx 40$.  The theoretical
distribution of noise power is exponential with a mean of
2, so the single trial probability that $Z_1^2>40$ by
chance is $2 \times 10^{-9}$.  Such a value is expected to arise
randomly in the $\sim 2 \times 10^9$ independent
trials that were performed, so it is not significant.
The expectation value of the intrinsic pulsed fraction
is $f_p = (1 + N_b/N_s)\sqrt{2Z_1^2/N}$,
where $N_b/N_s = 0.06/0.94$ is the ratio of background
to source counts, and
$N = N_s + N_b = 11,486$ is the total number of counts.
In this case, a sinusoidal
signal of $f_p = 9.0\%$ would have a 50\% probability of
giving $Z_1^2>40$.  However, taking into account the
noise, which may either increase or decrease the total power,
the 99\% upper limit on the $Z_1^2$ of a true
signal in this case is 75 \citep{gro75,vau94}, corresponding to
$f_p < 12.2\%$.  On the other hand, these peak power
values are somewhat overestimated because we oversampled
Fourier space.  Finally, then, we adopt an upper limit
of $f_p < 12\%$ for $P > 10$~ms in Table~\ref{ccos},
which is the lowest limit yet obtained for Cas~A over the range of
periods that is spanned by the known CCO pulsars.

\end{document}